\documentclass[prl,twocolumn,showpacs,amsmath,amssymb,superscriptaddress]{revtex4-1}
\usepackage{graphicx}
\usepackage{hyperref}
\usepackage{xcolor}
\usepackage{color}
\makeatother

\begin{document}

\title{Electric control of  Dirac quasiparticles by spin-orbit torque in an antiferromagnet}
\author{L. \v{S}mejkal}
\affiliation{Institute of Physics, Academy of Sciences of the Czech Republic, Cukrovarnick\'{a} 10, 162 53 Praha 6 Czech Republic}
\affiliation{Institut f\"ur Physik, Johannes Gutenberg Universit\"at Mainz, D-55099 Mainz, Germany}
\affiliation{Faculty of Mathematics and Physics, Charles University in Prague,
Ke Karlovu 3, 121 16 Prague 2, Czech Republic}
\author{J. \v{Z}elezn\'{y}}
\affiliation{Max Planck Institute for Chemical Physics of Solids, N\"{o}thnitzer Str. 40, D-01187 Dresden, Germany}
\affiliation{Institute of Physics, Academy of Sciences of the Czech Republic, Cukrovarnick\'{a} 10, 162 53 Praha 6 Czech Republic}
\author{J. Sinova}
\affiliation{Institut f\"ur Physik, Johannes Gutenberg Universit\"at Mainz, D-55099 Mainz, Germany}
\affiliation{Institute of Physics, Academy of Sciences of the Czech Republic, Cukrovarnick\'{a} 10, 162 53 Praha 6 Czech Republic}
\author{T. Jungwirth}
\affiliation{Institute of Physics, Academy of Sciences of the Czech Republic, Cukrovarnick\'{a} 10, 162 53 Praha 6 Czech Republic}
\affiliation{School of Physics and Astronomy, University of Nottingham, Nottingham NG7 2RD, United Kingdom}

\date{\today}

\begin{abstract}
Spin-orbitronics and Dirac quasiparticles are two fields of condensed matter physics initiated independently about a decade ago. Here we predict that Dirac quasiparticles can be controlled by the spin-orbit torque reorientation of the N\'eel vector in an antiferromagnet.  Using CuMnAs as an example, we formulate symmetry criteria allowing for the co-existence of Dirac quasiparticles and N\'eel spin-orbit torques. We identify the non-symmorphic crystal symmetry protection of Dirac band crossings whose  on and off switching is mediated by the N\'eel vector reorientation.  We predict that this concept, verified by minimal model and density functional calculations in the CuMnAs semimetal antiferromagnet, can lead to a topological metal-insulator transition driven by the N\'eel vector and to the corresponding topological anisotropic magnetoresistance. 
\end{abstract}
\maketitle

\twocolumngrid

2004 was the year when the spin Hall effect was observed in GaAs \cite{Kato2004d,Wunderlich2005,Sinova2015} and one-atom-thick flakes of graphene were isolated \cite{Novoselov2005,CastroNeto2009b}. The former discovery marked the dawn of the field of spin-orbitronics, in which the relativistic conversion between linear momentum and spin angular momentum of conducting electrons has provided new physical concepts for spintronics devices. These include the spin-orbit torque (SOT), which has opened the path to reliable and fast information writing in a ferromagnetic random access memory \cite{Miron2011b,Liu2012},  and also to efficient means of the electrical switching of an antiferromagnet (AF) by the N\'eel SOT \cite{Zelezny2014,Wadley2016}. Independently, the discovery of graphene initiated intense research of Dirac fermion quasiparticles in condensed matter systems. The field includes topological insulators, semimetals, or superconductors, which host a family of quasiparticles mimicking different flavors of fermions from relativistic particle physics \cite{Hasan2010a,Qi2011,Bansil2016}. More recently, novel phenomena have been discovered at the intersection of these two fields, such as the quantum spin Hall effect and the quantum anomalous Hall effect in non-magnetic and magnetic topological insulators  \cite{Kane2005a,Bernevig2006n,Roth2009,Chang2013,Chang2015,Bestwick2015}. Dirac quasiparticles, exhibiting a strong spin-momentum locking, are also considered for enhancing the efficiency of the SOT control of magnetic moments in ferromagnetic topological insulator hetero-structures \cite{Fan2014a}.  

In this Letter we close the loop of synergies between the fields of spin-orbitronics and Dirac quasiparticles by proposing a scheme for the electric control of Dirac band crossings  via the N\'eel SOT in AFs.  Our work addresses the outstanding problem of finding efficient means for controlling Dirac quasiparticles  by external fields which may provide the desired tools for the experimental research and future practical applications in microelectronics \cite{Dong2015}. On a specific example of the semimetal CuMnAs AF \cite{Wadley2016,Maca2012,Tang2016} we demonstrate that the N\'eel vector orientation is a suitable degree of freedom that can mediate on and off switching of the symmetry protection of Dirac band crossings. Based on this we also predict the topological metal-insulator transition (MIT) and the corresponding topological anisotropic magnetoresistance (AMR) in Dirac semimetal AFs.

\begin{figure}[t]
\centering
\includegraphics[width=0.5\textwidth]{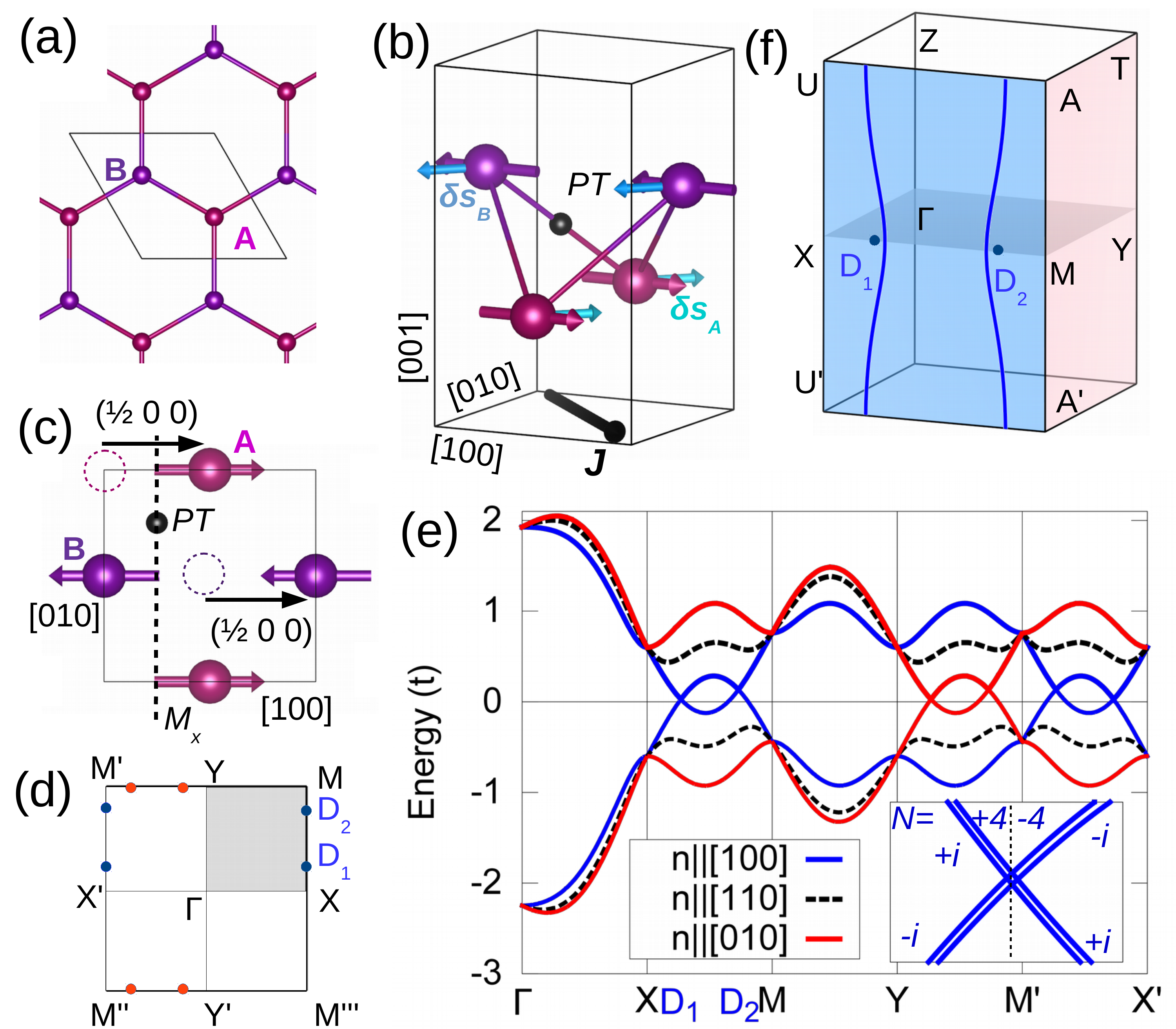}
\caption{(a) Graphene two-$A$,$B$-sublattice crystal. (b) Minimal two-$A$,$B$-sublattice AFM with the non-symmorphic symmetry extracted from CuMnAs. The two magnetic sublattices are connected by the $\cal{P}\cal{T}$ symmetry center marked by the black ball, and they are non-centrosymmetric. This allows for the nonzero staggered non-equilibrium spin-polarizations $\delta\textbf{s}_{A,B}$ induced by the current \textbf{J}, which facilitates the manipulation of the N\'eel vector. (c) Top view of our quasi-2D-AFM model highlighting the non-symmorphic glide mirror plane $\mathcal{G}_{x}$ (see text). (d) 2D BZ projection with the DPs positions - for \textbf{n}$\parallel$[100] along the MX axis (blue), while for \textbf{n}$\parallel$[010] along M$^\prime$Y axis (red). (e) Band dispersion of our minimal AFM model illustrating the control of the DPs and topological indexes of the  DP $D_{1}$ in the inset (for the sake of clarity the degenerate bands are slightly shifted). (f) 3D model BZ with the Dirac nodal lines (colors of the planes protected for a given N\'eel vector orientation correspond to (d,e)).} 
\label{Fig1}
\end{figure}

Dirac quasiparticles and the N\'eel SOT can co-exist because of the serendipitous overlap of the key symmetry requirements. We illustrate this on examples shown in Figs.~1(a),(b) of the graphene  lattice, representing the Dirac systems \cite{bernevig2013topological}, and the tetragonal CuMnAs crystal where the N\'eel SOT has been experimentally verified \cite{Wadley2016}: 

\noindent (i) The two-Mn-site primitive cell of CuMnAs favors band crossings in analogy with the two-C-site graphene lattice. 

\noindent (ii) In the paramagnetic phase, CuMnAs has time reversal ($\cal{T}$) and space inversion ($\cal{P}$) symmetries. It guarantees that each band is double-degenerate forming a Kramer's pair, in analogy to graphene. 
In the AF phase, this degeneracy is not lifted because the combined $\cal{PT}$ symmetry is preserved, although the $\cal{T}$ symmetry and the $\cal{P}$ symmetry are each broken  \cite{Herring1966,cracknell1975book,Chen2014,Tang2016}. This highlights antiferromagnetism as the favorable type of magnetic order for controlling Dirac quasiparticles. 

\noindent (iii) Finally, the combined $\cal{PT}$ symmetry  also allows for the efficient SOT reorientation of the N\'eel vector \cite{Zelezny2014,Zelezny2016}. Because the  $A$ and $B$ Mn-sites in the CuMnAs primitive cell are non-centrosymmetric inversion partners, a non-equilibrium spin-polarization $\delta \textbf{s}_{A,B}$  with opposite sign on the two sites is generated by an electrical current $\textbf{J}$ (see Fig.~1(b)) \cite{Zelezny2014}. This applies to both the paramagnetic and the AF phase above and below the N\'eel temperature of CuMnAs. Moreover, in the AF phase, the inversion partner $A$ and $B$ sites are occupied by oppositely oriented Mn magnetic moments (hence the combined $\cal{PT}$ symmetry). The current-induced non-equilibrium spin-polarization and the equilibrium AF moments are therefore both staggered and commensurate. In combination with the exchange interaction that couples them, the resulting current-induced N\'eel SOT  can efficiently reorient the N\'eel vector \cite{Zelezny2014,Wadley2016}.

An additional crystal symmetry is now needed to mediate the dependence of Dirac quasiparticles on the N\'eel vector orientation. In graphene there is no symmetry that protects the four-fold degeneracy of Dirac crossings of two Kramer's pair bands in the presence of spin-orbit coupling (SOC) \cite{Kane2005a}. Inspired by recent predictions of the symmetry protection of band-crossings in Dirac semimetals \cite{Wang2012g,Young2012,Young2015}, we identify non-symmorphic symmetries that can be turned on and off by reorienting the N\'eel vector in CuMnAs and by this can close and open a gap at the Dirac crossing. Recall that non-symmorphic space groups contain point group operations coupled with non-primitive lattice translations. 

We illustrate the concept first on a minimal model of the tetragonal CuMnAs AF, considering only the Mn atoms (with one orbital per atom) that form a stack of the crinkled quasi-2D square lattices shown in Figs.~1(b),(c). We first neglect the coupling between these quasi-2D planes; their distance is larger  than first and second nearest neighbor distances within the quasi-2D plane. The corresponding model Hamiltonian in the crystal momentum space,
\begin{eqnarray}
H_{\textbf{k}}=-2t\tau_{x}\cos\frac{k_{x}}{2}\cos\frac{k_{y}}{2}-t'\cos k_{x} \cos k_{y} +\nonumber \\ 
 \lambda \tau_{z}\left( \sigma_{y} \sin k_{x} - \sigma_{x} \sin k_{y} \right) + \tau_{z}J_n\boldsymbol\sigma\cdot\textbf{n}\,,
 \label{hamiltonian}
\end{eqnarray}
consists of the first nearest neighbor hopping $t$ (inter-sublattice $A-B$ hopping), the second nearest neighbor hopping $t'$ (intra-sublattice $A-A$ hopping), the second-neighbor SOC of strength $\lambda$ \cite{Kane2005a}, and the AF exchange coupling of strength $J_n$. The wavevector $k_{x(y)}$ is in units of the inverse lattice constant, $\textbf{n}$ is the N\'eel vector, and $\boldsymbol\tau$ and $\boldsymbol\sigma$ are Pauli matrices describing the crystal sublattice $A,B$ and spin degrees of freedom, respectively. We diagonalize $H$ analytically,
\begin{eqnarray}
E_{\textbf{k}\pm}=-t'\cos k_{x} \cos k_{y} \pm  [ 4t^{2}\cos^{2}\frac{k_{x}}{2}\cos^{2}\frac{k_{y}}{2}+\nonumber \\
\hspace*{-.2cm} (J_nn_x-\lambda \sin k_{y})^{2}
+(J_nn_y+\lambda \sin k_{x})^{2}+J_n^{2}n_{z}^{2}] ^{1/2}\,,
\label{dispersion}
\end{eqnarray}
and plot in Fig.~1(e) the resulting bands measured from the Fermi level for $\lambda=0.8t, J_n=0.6t$, and $t^\prime=0.08t$. For the N\'eel vector \textbf{n}$\parallel$[100] we found two Dirac points (DPs) $D_{1}$, and $D_{2}$ in the first Brillouin zone (BZ) along the  MX axis at wave-vectors $\textbf{D}_{1}=(\pi,\arcsin \frac{J_n}{\lambda})$ and $\textbf{D}_{2}=(\pi,\pi-\arcsin \frac{J_n}{\lambda})$, as shown in Figs.~1(d-f). 

For the minimal quasi-2D model of the CuMnAs AF we can now show explicitly that the DPs are protected by a non-symmorphic, glide mirror plane symmetry \cite{Young2015}, $\mathcal{G}_{x}=\left\lbrace M_{x} \vert \frac{1}{2} 0 0 \right\rbrace$. It combines the mirror symmetry $M_{x}$  along the (100)-plane with the half-primitive cell translation along the [100] axis (see Fig.~1(c)) and has eigenvalues $g_{\pm}=\pm i$. The four-fold degenerate DP originates from a crossing of two Kramer's pairs where the two bands in each pair are degenerate due to the $\mathcal{PT}$ symmetry. Hybridization between the pairs is prohibited, i.e. the crossing is protected by $\mathcal{G}_{x}$, when the following conditions are met: (i) The crossing occurs at the BZ sub-manifold invariant under $\mathcal{G}_{x}$. This is fulfilled in $k_{x}=0,\pm\pi$ planes. (ii) The two bands forming a given Kramer's pair,  with the corresponding wave-functions $\psi_{\textbf{k}}$ and $\mathcal{PT} \psi_{\textbf{k}}$, can be assigned the same eigenvalue of $\mathcal{G}_{x}$. From the commutation relation of $\mathcal{G}_{x}$ and $\mathcal{PT}$ we obtain that this condition is fulfilled only at the BZ sub-manifold $k_{x}=\pm\pi$ (cf. Figs.~1(d),(e)). (iii) One Kramer's pair corresponds to one eigenvalue and the other Kramer's pair to the opposite eigenvalue of $\mathcal{G}_{x}$. This can be verified by employing the $k \cdot p$ perturbation theory.  Around, e.g., the $D_{1}$ point  in the $k_{x}=\pi$ plane we obtain, $E_{\textbf{D}_{1}+k_{y},\pm}=\pm \hbar v_{F,y} k_{y}$ with the two Kramer's pairs fulfilling, $\mathcal{G}_{x}\psi_{\textbf{k}\pm}=\mathcal{G}_{x}\mathcal{PT}\psi_{\textbf{k}\pm}=\mp i \psi_{\textbf{k}\pm}$. This is highlighted in the inset of Fig.~1(e). 

Because of the combined $\mathcal{PT}$ symmetry we can define the topological index of our DPs analogously to the paramagnetic Dirac semimetals \cite{yang2016}. The topological index $N(k_{y})$ at the crystal momentum $k_{y}$, invariant under $\mathcal{G}_{x}$, is given by: $N(k_{y})=\left[N^{C}_{+i}(k_{y})-N^{V}_{+i}(k_{y}) \right]-\left[N^{C}_{-i}(k_{y})-N^{V}_{-i}(k_{y}) \right]$. Here $N^{C(V)}_{\pm i}(k_{y})$ is the number of conduction (valence) bands  at $k_y$ with the eigenvalue   $g_{\pm}=\pm i$. An integer value and a discontinuity  of the topological index when crossing the DP in our model is highlighted in the inset of Fig.~1(e). The corresponding topological charge at e.g. the DP $D_{1}$, obtained by approaching $\textbf{D}_{1}$ from left and right \cite{yang2016},  is $Q\equiv\left[N(\textbf{D}_{1}+\delta)-N(\textbf{D}_{1}-\delta) \right]/8=-1$.

Following the symmetry analysis of N\'eel SOTs in Refs.~\cite{Zelezny2014,Zelezny2016}, we obtain for our CuMnAs model that the lowest-order (\textbf{n}-independent) component of $\delta \textbf{s}_{A,B}$
is staggered, i.e. can generate an efficient field-like SOT. The field allows for the rotation of  \textbf{n} in the (001)-plane in the direction perpendicular to the applied in-plane current. In Figs.~1(d),(e) we show that for \textbf{n}$\parallel$[010] the DPs move to the M$^\prime$Y axis. 
They are now protected by the $\mathcal{G}_{y}=\left\lbrace M_{y} \vert 0\frac{1}{2} 0 \right\rbrace$ symmetry, as expected for the square quasi-2D lattice. 
\begin{figure}[h]
\centering
\includegraphics[width=0.5\textwidth]{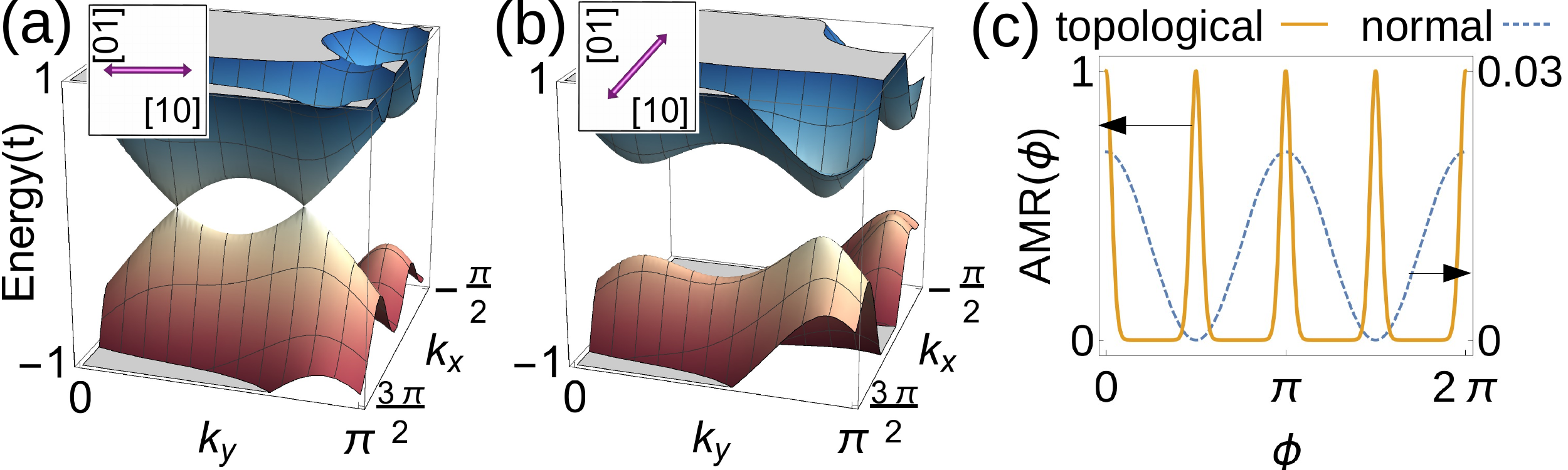}
\caption{Topological MIT in our minimal model driven by N\'eel vector reorientation from  (a) [100] to (b) [110] by the N\'eel SOT. (c) Schematics of the corresponding angular dependence of the topological AMR contrasted to the normal AMR.} 
\label{Fig2}
\end{figure}

At intermediate in-plane angles, no DP protecting symmetry  remains and the entire spectrum is gapped (see the \textbf{n}$\parallel$[110] bands in Fig.~1(e)). As highlighted on the full spectra of the minimal quasi-2D model in Fig.~2(a),(b), this leads to the topological MIT driven by the N\'eel vector reorientation. The band-gap at the DP is a continuous function of the in-plane N\'eel vector angle, $\Delta(\textbf{D}_{1})\sim J_{n}\sqrt{1-\cos(\phi)}$, where $\phi$ is measured from the [100] axis. The transport counterpart of the MIT is the topological AMR which we define as, AMR$\equiv[\sigma(\phi)-\sigma_{\text{min}}]/\sigma_{\text{max}}$. Here  $\sigma(\phi)$ is the $\phi$-dependent conductivity with current along the [100] axis and $\sigma_{\text{min(max)}}$ referring to the conductivity minimum (maximum). The topological AMR in  our  Dirac semimetal AF is schematically illustrated in Fig.~2(c). High AMR values correspond to $\phi=0(\pi/2)$ with the closed gap of the DPs at MX (M$^\prime$Y). 
For comparison, we illustrate in Fig.~2(c) a characteristic harmonic angular dependence of the conventional  AMR in a normal magnetic metal \cite{Marti2014}. 

We conclude the discussion of the minimal model of the CuMnAs AF by taking into account the coupling between the quasi-2D planes. The coupling leads to the following renormalization of the model Hamiltonian \eqref{hamiltonian}: 
$ 
2t\tau_{x} \stackrel{3D}{\rightarrow} (2t+t_{z}\cos k_{z})\tau_{x}+t_{z}\sin k_{z}\tau_{y},  
t'(\cos k_x + \cos k_y) \stackrel{3D}{\rightarrow} t'(\cos k_x + \cos k_y)+t_{z}'\cos k_{z}$, and $
\lambda \stackrel{3D}{\rightarrow}  \lambda - \lambda_{z} \cos k_z.
$
As a result, the $\mathcal{G}_{x}$ ($\mathcal{G}_{y}$) protected DPs in  2D transform into protected nodal lines in 3D. For example, $\textbf{D}_{1}\stackrel{3D}{\rightarrow}(\pi,\arcsin \frac{J_{n}}{\lambda-\lambda_{z}\cos k_{z}},k_{z})$ gives an open nodal line for $\lambda_{z} < \lambda/2$, as shown in Fig.~1(f)  for $\lambda_{z}=0.2t$. Note that in our Dirac AF, the nodal lines are dispersive in contrast to the paramagnetic $J_{n}=0$ model \cite{yang2016}.
\begin{figure}[t]
\centering
\includegraphics[width=0.5\textwidth]{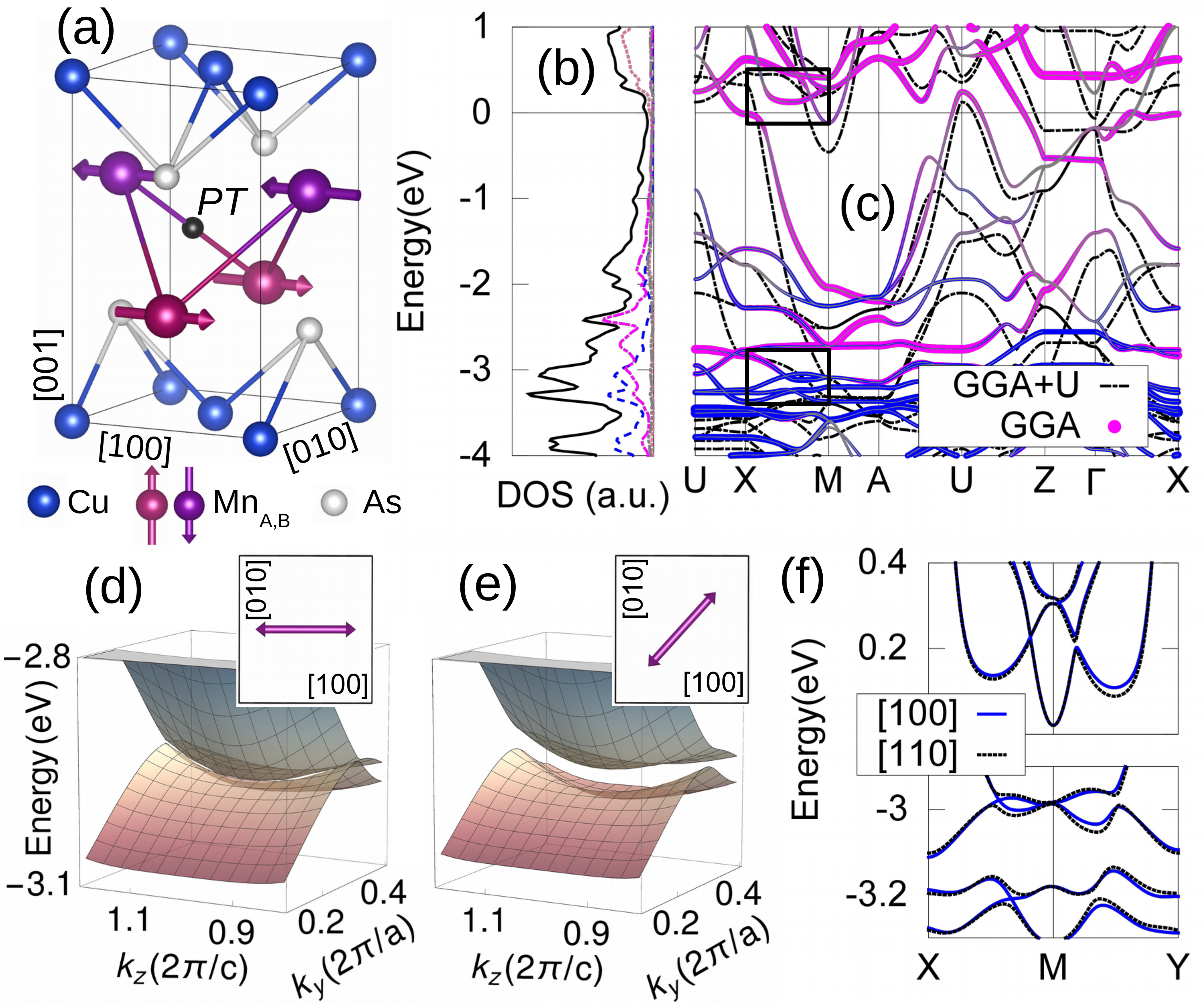}
\caption{(a) Crystallographic and magnetic structure of the tetragonal CuMnAs. Atom-resolved (b) density of states with semi-metallic pseudogap, and (c) band structure without SOC within GGA. GGA+U shows DPs position shifts. Colors correspond to the atomic colors in (a). Electric control by the N\'eel SOT of the 3D band dispersion around nodal line along the $k_{x}=\pi$ BZ sub-manifold calculated by GGA+SOC, which is (d) protected for \textbf{n}$\vert$[100] by glide mirror plane, (e) gapped for \textbf{n}$\vert$[110]. $a=b \neq c$ are the lattice constants.  (f) Cut along the XMY line through the nodal lines at different energies.} 
\label{Fig3}
\end{figure}

We now verify all observations made in the minimal model by performing full-potential relativistic {\em ab initio} calculations as implemented in \cal{FLEUR} and \cal{ELK} packages \cite{codes}. 
The exchange correlation potential is parametrized by the Perdew-Burke-Ernzerhof generalized gradient approximation (GGA) \cite{perdew1996,perdew1997}. The full crystal of tetragonal CuMnAs, including also the Cu and As atoms, is shown in Fig.~3(a) \cite{Wadley2013,Wadley2015a}. Results without SOC are summarized in Figs.~3(b),(c). They show the semimetallic  character with the dip in the density of states near the Fermi level and numerous band crossings. Note that their position is sensitive to the computational details; as an illustration we plot in Fig.~3(c) shifted bands obtained in the GGA+U approximation with the correlation potential $U=3$~eV. 
When SOC is included in the {\em ab initio} calculations and \textbf{n}$\parallel$[100], protected nodal lines are obtained in the $k_{x}=\pm\pi$ planes, as illustrated in Figs.~3(d). The nodal lines have the open geometry (cf. Fig.~1(f)). The protection is due to the $\mathcal{G}_{x}$ symmetry, also in agreement with the minimal model. Instead of assigning the $\mathcal{G}_{x}$  eigenvalues in the complex  {\em ab initio} band structure, we verify this by excluding all other relevant symmetries as the origin of the protection. For \textbf{n} $\parallel$[100], the space group $P4/nmm$ of the tetragonal CuMnAs lattice reduces to eight symmetry elements: Identity, non-symmorphic glide planes $\mathcal{G}_{x}$, and $\mathcal{G}_{z}=\left\lbrace M_{z} \vert \frac{1}{2} \frac{1}{2} 0\right\rbrace$, screw-axis  $\mathcal{S}_{y}$=$\left\lbrace C_{2y} \vert 0 \frac{1}{2} 0\right\rbrace$, and four $\cal{PT}$ conjugated symmetries. By rotating the N\'eel vector to \textbf{n}$\parallel$[110] and \textbf{n}$\parallel$[101], $\mathcal{G}_{z}$ and $\mathcal{S}_{y}$ remain symmetries of the AF crystal, respectively. In both cases, however, the nodal lines become gapped, as illustrated in Fig.~3(e), excluding the protection by these symmetries. Note that the $\mathcal{G}_{x}$ protection makes our tetragonal CuMnAs AF distinct from the earlier identified non-symmorphic protection in paramagnetic ZrSiS \cite{Schoop2016}. 
\begin{figure}[t]
\centering
\includegraphics[width=0.5\textwidth]{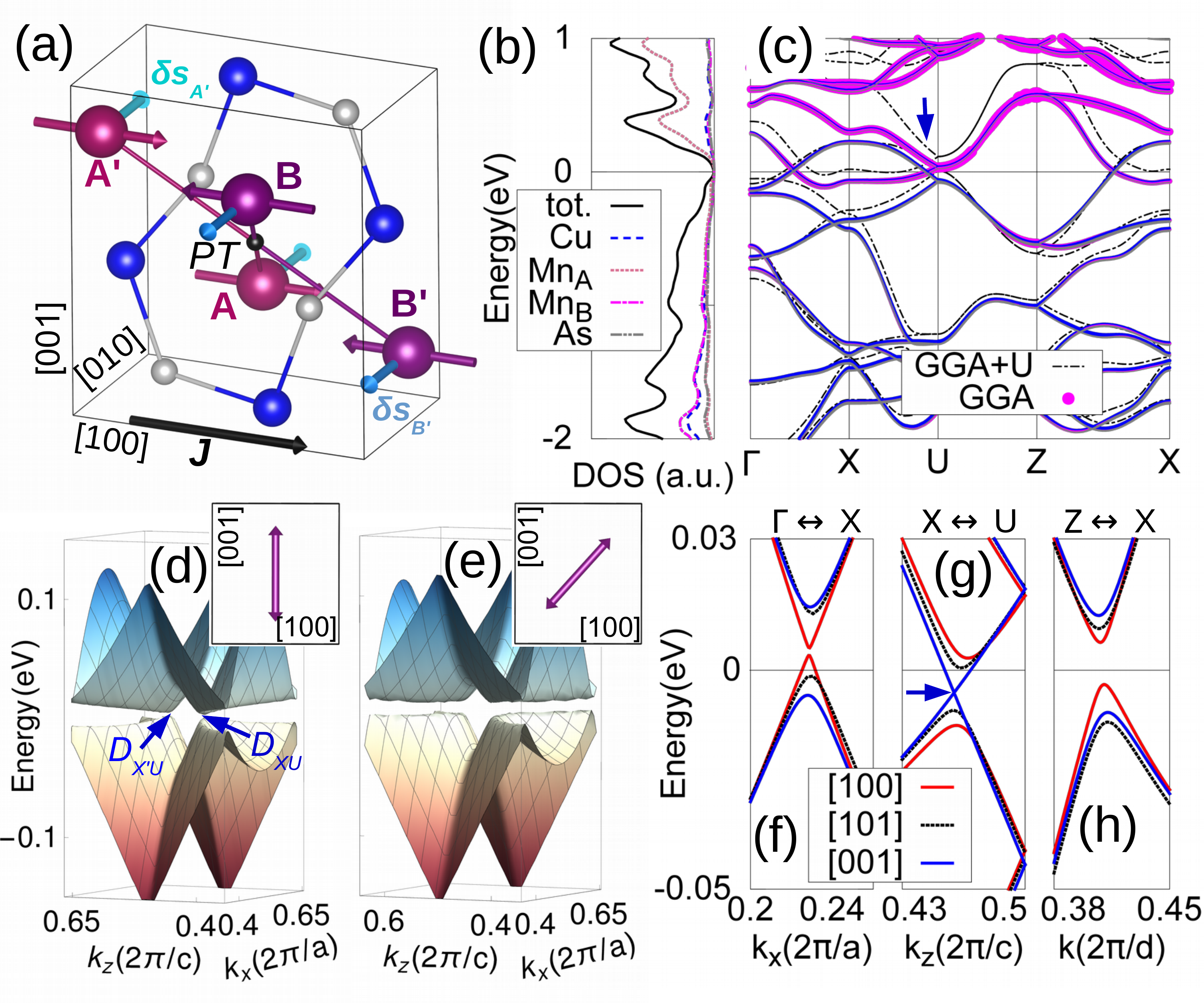}
\caption{(a) Crystallographic and magnetic structure of the orthorhombic CuMnAs with N\'eel SOT spin-polarization $\delta$\textbf{s} for the current \textbf{J}$\parallel$[100]. Atom-resolved (b) point-semimetal density of states, and (c) band structure without SOC within GGA. GGA+U shows DPs position shifts. (d-e) Topological MIT.  Manipulation of the Dirac fermions along the (f) $\Gamma$X, (g) XU, and (h) ZX axis (units $d=\sqrt{a^{2}+c^{2}}$ with $a \neq c$ being the lattice constants) by the N\'eel SOT from GGA+SOC calculations reveals: topological (\textbf{n}$\parallel$[001]), and "trivial" Dirac semimetal (\textbf{n}$\parallel$[100]), and semiconductor (\textbf{n}$\parallel$[101]). 
}
\label{Fig4}
\end{figure}

The field-like N\'eel SOT in the full tetragonal crystal of CuMnAs has the same symmetry as in the minimal model and, therefore, allows for the current-induced rotation of the N\'eel vector \cite{Wadley2016,Zelezny2016}. This opens the prospect of electric control of Dirac crossings in an experimentally relevant AF material. However, the tetragonal CuMnAs is not optimal for observing the corresponding topological MIT due to other non-Dirac bands present around the Fermi level (see Fig.~3(c)). These can be removed e.g. by lowering the lattice symmetry from tetragonal to orthorhombic \cite{Maca2012,Tang2016}, as we now discuss in the remaining paragraphs. 

The non-symmorphic  $Pnma$ primitive cell of the orthorhombic CuMnAs \cite{mundelein} is shown in Fig.~4(a). 
It has four Mn sites consisting of the two inversion-partner pairs $A$-$B$ and $A^\prime$-$B^\prime$. From the symmetry analysis  of the current-induced spin-polarizations \cite{Zelezny2016} generated locally at these four sites we obtain that they contain components which are commensurate with the AF order: $A$ and $A^\prime$ sites with one sign of the current-induced spin-polarizations belong to one AF spin-sublattice and $B$ and $B^\prime$ sites with the opposite sign of the current-induced spin-polarizations belong to the opposite AF spin-sublattice. This makes the N\'eel SOT efficient for reorienting AF moments in orthorhombic CuMnAs. 

GGA electronic structure calculations without SOC are shown in Fig.~4(b),(c). Consistent with earlier reports \cite{Maca2012,Tang2016}, the density of  states vanishes at the Fermi level and we now discuss the properties of the three Fermi level DPs seen in Fig~4(b),(c). Without SOC they are part of an ungapped nodal line in the  $k_{y}=0$ plane \cite{Tang2016}. In the presence of SOC and for \textbf{n}$\parallel[001]$, the DPs along the $\Gamma$X and ZX axes become gapped. The gap opening applies to the entire nodal line,  except for the DP along the XU axis (and also X$^\prime$U), as shown in Figs.~4(d),(f)-(h). Using the same method as in the {\em ab initio} calculations for the tetragonal CuMnAs, we identified that the XU DP protection is due to the screw-axis symmetry $S_{z}=\left\lbrace C_{2z}\vert \frac{1}{2} 0 \frac{1}{2} \right\rbrace$ \cite{Tang2016}. The corresponding state at \textbf{n}$\parallel$[001] is then a topological AF Dirac semimetal with the topological charge of the XU DP $Q=+1$. For \textbf{n}$\parallel$[101], all DPs (the entire nodal line) are gapped and the system becomes an AF semiconductor, as seen in Figs.~4(e)-(h). Finally, for \textbf{n}$\parallel$[100], the spin-orbit gap is nearly but not fully closed at the $\Gamma$X DP, as shown in Fig.~4(f). This "trivial" AF Dirac semimetal phase is reminiscent of graphene. Since the DPs 
can appear at the Fermi level (see also the comparison of GGA and GGA+U calculations in Fig.~4(b)), orthorhombic CuMnAs represents a realistic material candidate for observing  the topological MIT and AMR driven by the 
N\'eel vector reorientation.

\begin{acknowledgments}
We acknowledge support from the Grant Agency of the Charles University no. 280815 and of the Czech Republic no. 14-37427, Humbolt foundation, EU ERC Synergy Grant No. 610115, and the Transregional Collaborative Research Center (SFB/TRR) 173 SPIN+X. Access to computing and storage facilities owned by parties and projects contributing to the National Grid Infrastructure MetaCentrum provided under the programme "Projects of Projects of Large Research, Development, and Innovations Infrastructures" (CESNET LM2015042), is greatly appreciated.
\end{acknowledgments}

%

\end{document}